\documentclass[10pt,conference]{IEEEtran}
\IEEEoverridecommandlockouts
\usepackage{cite}
\usepackage{amsmath,amssymb,amsfonts}
\usepackage{mathtools} 
\usepackage{graphicx}
\usepackage{subcaption}
\usepackage{textcomp}
\usepackage{xcolor}
\usepackage{tabularx}
\usepackage{booktabs}
\usepackage{float}
\usepackage{algpseudocode}
\usepackage[english]{babel}
\usepackage{paralist}
\usepackage{array}
\usepackage{amsmath,stackrel}
\usepackage{multicol}
\usepackage{verbatim}
\usepackage{enumerate}              
\usepackage{dsfont}
\usepackage{psfrag}
\usepackage{algorithm}
\usepackage[nolist]{acronym}
\usepackage[utf8]{inputenc}
\usepackage{tikz}
\usetikzlibrary{external}
\usepackage{collcell}
\usepackage{pgfplots}
\pgfplotsset{compat=1.18}
\usepackage{hhline}
\usepackage{colortbl}
\usepackage{multido}
\usepackage{multirow}
\usepackage{dblfloatfix}    
\usepackage{etoolbox}
\usepackage{soul}
\usepackage{eucal}
\usepackage{cuted}
\usepackage[font=footnotesize]{caption}
\makeatletter
\let\oldfootnote\footnote
\def\footnote{\@ifstar\footnote@star\footnote@nostar}
\def\footnote@star#1{{\let\thefootnote\relax\footnotetext{#1}}}
\def\footnote@nostar{\oldfootnote}
\makeatother
\begin{acronym}
\acro{AWGN}{additive white Gaussian noise}
\acro{BER}{bit error rate}
\acro{BS}{base station}
\acro{CP}{cyclic prefix}
\acro{CRLB}{Cramér-Rao lower bound}
\acro{FMCW}{frequency modulated continuous wave}
\acro{ICI}{inter-carrier interference}
\acro{IoT}{Internet of Things}
\acro{ISAC}{integrated sensing and communication}
\acro{ISFFT}{inverse symplectic finite Fourier transform}
\acro{ISI}{inter-symbol interference}
\acro{JSC}{joint sensing and communication}
\acro{LoS}{line of sight}
\acro{MIMO}{multiple-input multiple-output}
\acro{ML}{maximum likelihood}
\acro{OFDM}{orthogonal frequency-division multiplexing}
\acro{OTFS}{Orthogonal time frequency space}
\acro{PER}{packet error rate}
\acro{QAM}{quadrature amplitude modulation}
\acro{QPSK}{quadrature phase shift keying}
\acro{RMSE}{root mean square error}
\acro{SFFT}{symplectic finite Fourier transform}
\acro{SNR}{signal-to-noise ratio}
\acro{UE}{user equipment}
\acro{V2V}{vehicle-to-vehicle}
\acro{V2I}{vehicle-to-infrastructure}
\acro{V2X}{vehicular-to-everything}
\acro{5G}{fifth generation}
\end{acronym}

\def\BibTeX{{\rm B\kern-.05em{\sc i\kern-.025em b}\kern-.08em
    T\kern-.1667em\lower.7ex\hbox{E}\kern-.125emX}}
\begin{document}


\title{Performance Analysis of a Low-Complexity OTFS Integrated Sensing and Communication System
\thanks{This work has been accepted for publication in \emph{Proc. IEEE VTC2023-Fall.}}
\thanks{\copyright{ 2023 IEEE}. Personal use of this material is permitted. Permission from IEEE must be obtained for all other uses, in any current or future media, including reprinting/republishing this material for advertising or promotional purposes, creating new collective works, for resale or redistribution to servers or lists, or reuse of any copyrighted component of this work in other works. }
\thanks{This work was supported by the European Union under the Italian National Recovery and Resilience Plan (NRRP) of NextGenerationEU, partnership on ``Telecommunications of the Future'' (PE00000001 - program ``RESTART'').}
}
\author{\IEEEauthorblockN{
Tommaso~Bacchielli, 
Lorenzo~Pucci,
Enrico~Paolini, 
and Andrea~Giorgetti
}
\medskip
\IEEEauthorblockA{Wireless Communications Laboratory, CNIT, DEI, University of Bologna, Italy\\ 
Email: \{tommaso.bacchielli2, lorenzo.pucci3, e.paolini, andrea.giorgetti\}@unibo.it\\
}

}

\maketitle

\begin{abstract}
This work proposes a low-complexity estimation approach for an orthogonal time frequency space (OTFS)-based \ac{ISAC} system.
In particular, we first define four low-dimensional matrices used to compute the channel matrix through simple algebraic manipulations. Secondly, we establish an analytical criterion, independent of system parameters, to identify the most informative elements within these derived matrices, leveraging the properties of the \emph{Dirichlet kernel}. This allows the distilling of such matrices, keeping only those entries that are essential for detection, resulting in an efficient, low-complexity implementation of the sensing receiver.
Numerical results, which refer to a vehicular scenario, demonstrate that the proposed approximation technique effectively preserves the sensing performance, evaluated in terms of \ac{RMSE} of the range and velocity estimation, while concurrently reducing the computational effort enormously.
\end{abstract}


\acresetall


\section{Introduction}

The need to provide future mobile networks with \ac{ISAC} capabilities has emerged recently. The sensing capability consists of the detection and localization of non-collaborative objects, a feature typical of radar systems, performed by sharing the hardware and the physical layer with the communication part of the system. Harnessing the potential of such systems paves the way for ubiquitous and pervasive sensing, ushering in a radically new paradigm commonly referred to as perceptive networks \cite{LiuetalJSAC22,PucPaoGio:J22,ISAC_Survey2023}.
\ac{OTFS}, a two-dimensional modulation technique performed in the delay-Doppler domain, has recently gained much interest in both communication and sensing applications. In fact, in addition to the higher spectral efficiency with respect to \ac{OFDM} modulation due to the absence of the \ac{CP} and the robustness in high-mobility scenarios \cite{Hadani2017}, this modulation has demonstrated notable suitability in \ac{ISAC} scenarios \cite{Gaudio2020,Mohammed2022,Yuan2023,Dehkordi2023}.
Unfortunately, \ac{OTFS} is also remarkably complex, making it challenging to implement efficiently \cite{Viterbo2021,Viterbo2022}. This problem is even more exacerbated in \ac{ISAC} systems where sensing parameters must be estimated quickly to support real-time tracking \cite{FavMatPuc:C23,FavMatPuc:C22}. 

At the forefront, various approaches in the literature aim to reduce and manage the complexity of \ac{OTFS}-based \ac{ISAC} systems \cite{OTFS-FMCW,low-complexity_param, On_OTFS_using_Zak}. In particular, in \cite{OTFS-FMCW}, a mixed technique that couples the \ac{FMCW} signal, suffering from low data rate, with \ac{OTFS} waveform, which suffers from high sensing complexity, is proposed to achieve high data rates for communication while ensuring a low-complexity radar receiver. In \cite{On_OTFS_using_Zak}, an efficient \ac{OTFS} implementation is presented using the discrete Zak transform. This approach simplifies the derivation and the analysis of the input-output relation of the time-frequency dispersive channel in the delay-Doppler domain.

Despite attempts to reduce and simplify the implementation complexity of \ac{OTFS} receivers, current solutions are not yet effective, especially for \ac{ISAC} applications, where quick estimation of target parameters is a strict requirement for real-time tracking in complex environments.

Therefore, this paper proposes a low-complexity implementation of the \ac{ML} estimator, used to infer range and velocity, through an original approximation criterion developed explicitly for the channel matrix expression proposed in \cite{Gaudio2020}. In particular, we first compute the channel matrix in \cite{Gaudio2020} via four low-dimensional matrices. Then, a criterion for identifying the most informative elements (for sensing purposes) of each of these matrices is derived based on an in-depth analysis of the \emph{Dirichlet kernel}. Starting from this, a general approximation criterion independent of system parameters is derived and applied to the four matrices to reduce the sensing system's computational complexity. We then assess the impact of this approximation technique in terms of the \ac{RMSE} for range and velocity estimation.

In this paper, we adopt the following notation: capital boldface letters for matrices, lowercase bold letters for vectors, $(\cdot)^\mathrm{H}$ for conjugate transpose of a vector/matrix, and $|\cdot|$, $\lceil \cdot \rceil$, $\lfloor \cdot \rfloor$ for absolute value, ceiling and floor functions, respectively. Also, $(\cdot)^\ast$, $\mathbb{E}[\cdot]$, $\mathrm{var}[\cdot]$ for conjugate, mean value and variance operator.  
Additionally, $\text{frac}(x) = x - \lfloor x \rfloor$ represents the fractional (or decimal) part of a non-negative real number $x$.

The paper is organized as follows. In Section~\ref{sec:System_model}, the system model is described, while Section~\ref{sec:Sensing_estimation} presents the ML estimator and the \ac{CRLB}. In Section~\ref{sec:Cross-talk_computation} and Section~\ref{sec:Cross-talk_approx}, we provide some considerations on the channel matrix in \cite{Gaudio2020} and describe the approximation technique, respectively. The numerical results are presented in Section~\ref{sec:Num_Res}, while Section~\ref{sec:Conclusion} concludes the paper.


\section{System Model} \label{sec:System_model}


\subsection{Physical Model} \label{sec:physic-mod}

This work considers a monostatic \ac{ISAC} \ac{OTFS}-based system. In particular, a monostatic \ac{ISAC} transceiver, with full-duplex capabilities, is used to jointly communicate with \acp{UE} and simultaneously sense the environment by collecting signals backscattered from targets. By acquiring these signals, the monostatic \ac{ISAC} system can estimate the distance and radial velocity of nearby targets.

Similar to \ac{OFDM}, \ac{OTFS} is a multicarrier modulation scheme that operates at a given carrier frequency $f_\mathrm{c}$, with a bandwidth $B=M \Delta f$, where $M$ is the number of subcarriers and $\Delta f=1/T$ is the subcarrier spacing, with $T$ the symbol time. Like \ac{OFDM}, the relationship between $\Delta f$ and $T$ is chosen to avoid \ac{ICI}. However, unlike \ac{OFDM}, in \ac{OTFS} systems, it is possible to avoid the use of a cyclic prefix, needed to prevent \ac{ISI}, by properly choosing the modulating and demodulating pulses \cite{Hadani2017}.
\begin{figure}
    \centering
    \includegraphics[width=.97\columnwidth]{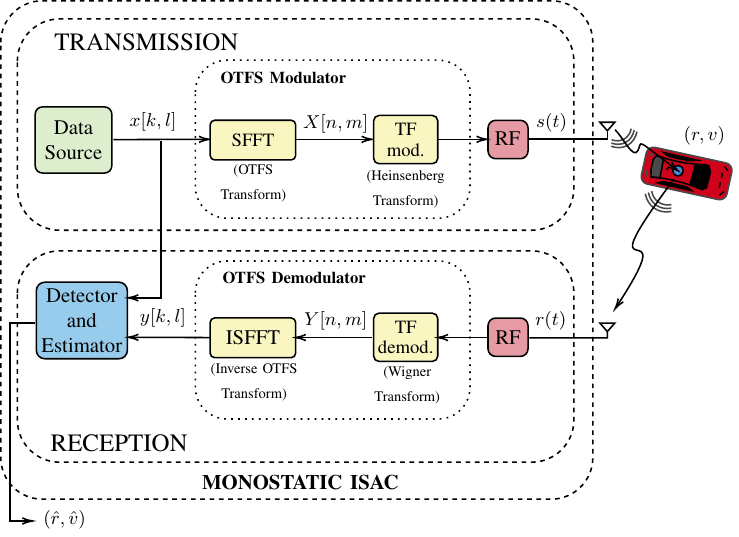}
    \caption{Signal-processing chain of the considered ISAC OTFS-based system.}
    \label{fig:ISAC_OTFS_system_model}
\end{figure}
Considering a generic scenario with $P$ point-like targets, each moving with a relative radial velocity $v_p$ at a distance $r_p$ from the monostatic transceiver, and considering \ac{LoS} propagation conditions between the transceiver and the targets, the time-varying channel impulse response of the radar system can be written as
\begin{equation} \label{eqn:h_channel}
    h(t,\tau) = \sum_{p=0}^{P-1} h_p \delta(\tau-\tau_p) e^{j2\pi f_{\mathrm{D},p} t}
\end{equation}
where $h_p$ is the complex channel gain, $f_{\mathrm{D},p}=\frac{2v_p f_\mathrm{c}}{c}$ the Doppler shift and $\tau_p=\frac{2r_p}{c}$ the round-trip delay of the signal backscattered by the target $p$.

The time-continuous transmitted signal $s(t)$ can be written as
\begin{equation} \label{eqn:s}
    s(t) = \sum_{n=0}^{N-1} \sum_{m=0}^{M-1} X[n,m] g_{\mathrm{tx}}(t-nT) e^{j2\pi m\Delta f (t-nT)}
\end{equation}
where $g_\mathrm{tx}$ is the pulse shape, while $X[n,m] \in \mathbb{C}$ represents a generic modulation symbol, taken from a complex modulation alphabet, and defined in the time-frequency domain symbol grid at the output of the pre-processing block of \ac{OTFS} modulator, as shown in Fig.~\ref{fig:ISAC_OTFS_system_model}. Symbols are arranged in the $N \times M$ grid $\mathbf{\Lambda}$, where $N$ is the number of time slots.\\
The so-called pre-processing operation consists of a \ac{SFFT} that maps the symbols $x[k,l]$, generated by the transmitter in the delay-Doppler domain and belonging to the grid $\mathbf{\Lambda}^\bot$ of $M \times N$ dimension, to the time-frequency domain symbols $X[n,m]$, as
\begin{equation} \label{eqn:X_JSC}
    X[n,m] = \sum_{k=0}^{M-1} \sum_{l=0}^{N-1} x[k,l] e^{-j2\pi(\frac{mk}{M}-\frac{nl}{N})}
\end{equation}
with $n,l=0,...,N-1$ and $m,k=0,...,M-1$.

Starting from \eqref{eqn:h_channel} and \eqref{eqn:s}, and neglecting for the moment the noise introduced by the communication channel, the continuous-time received signal $r(t)$ is given by
\begin{align} \label{eqn:r}
    r(t) &= \int h(t,\tau) s(t-\tau) d\tau \nonumber\\ &=\sum_{p=0}^{P-1} h_p s(t-\tau_p) e^{j2\pi f_{\mathrm{D},p} t}.
\end{align}
The received signal $r(t)$ is first given as an input to a matched filter, which performs cross-correlation between $r(t)$ and a demodulating pulse $g_{\mathrm{rx}}$, and then discretizes the output signal in both the time and frequency domains.\\
As a result, first, the cross-ambiguity function $A_{g_{\mathrm{rx}},r}$, defined in the time-frequency domain, is obtained, as
\begin{align} \label{eqn:Y_hat_OTFS_full} 
    Y(t,f) & = A_{g_{\mathrm{rx}},r}(\tau,f_{\mathrm{D}})|_{\tau=t,f_{\mathrm{D}}=f}\nonumber\\
    & = \int r(t') g_{\mathrm{rx}}^\ast(t'-t) e^{-j2\pi f(t-t')} dt'.
\end{align}
Then, after a few mathematical manipulations, by sampling \eqref{eqn:Y_hat_OTFS_full} at times $t=nT$ and frequencies $f=m \Delta f$, a $N \times M$ symbol grid in the time-frequency domain is obtained, whose generic $(n,m)$ element is given by
\begin{align} \label{eqn:Y_symbol} 
    Y[n,m] & =  Y(t,f)|_{t=nT,f=m\Delta f}\nonumber\\
    & =  \sum_{n'=0}^{N-1} \sum_{m'=0}^{M-1} X[n',m'] H_{n,m}[n',m']
\end{align}
\noindent where, letting $h'_p \triangleq h_p e^{j2\pi f_{\mathrm{D},p} \tau_p}$, $H_{n,m}[n',m']$ is defined as
\begin{align}
    H_{n,m} & [n',m']
    \triangleq \sum _{p=0}^{P-1} h_{p}'e^{j2\pi n'Tf_{\mathrm{D},p}}e^{-j2\pi m\Delta f\tau _{p}} \nonumber \\
    &\times A_{ g_{\mathrm {rx}}, g_{\mathrm {tx}}}\left ({\left ({n-n'}\right)T-\tau _{p},\left ({m-m'}\right)\Delta f-f_{\mathrm{D},p}}\right).
\end{align}
\begin{figure*}[!t]
\begin{equation} \label{eqn:cross-talk_official}
{\scriptsize{%
        \mathbf{\Psi}_{l,l'}^{p}[k,k'] \approx \frac{1}{NM} \frac{1-e^{j2\pi(l'-l+f_{\mathrm{D},p} NT)}}{1-e^{j2\pi\frac{(l'-l+f_{\mathrm{D},p} NT)}{N}}} \frac{1-e^{j2\pi(k'-k+\tau_p M\Delta f)}}{1-e^{j2\pi\frac{(k'-k+\tau_p M\Delta f)}{M}}} e^{j2\pi f_{\mathrm{D},p}\frac{k'}{M\Delta f}}
         \begin{cases} 
            1 & k'=0,...,(M-1-k_{\tau_p})=k'_{\mathrm{I}\mathrm{C}\mathrm{I}}\\
            e^{-j2\pi(\frac{l'}{N}+f_{\mathrm{D},p} T)} & k'=(M-k_{\tau_p}),...,(M-1)=k'_{\mathrm{I}\mathrm{S}\mathrm{I}}.
        \end{cases}
}}
\end{equation}
\hrule
\end{figure*}

Lastly, the demodulated symbols, back in the delay-Doppler domain, are obtained after the post-processing block by performing the \ac{ISFFT} on the symbols in \eqref{eqn:Y_symbol}, as
\begin{eqnarray} \label{eqn:y_hat_ISAC} \nonumber
    y[k,l] & = & \frac{1}{NM} \sum_{n=0}^{N-1} \sum_{m=0}^{M-1} Y[n,m] e^{j2\pi(\frac{mk}{M}-\frac{nl}{N})}\\
    & = & \sum_{k'=0}^{M-1} \sum_{l'=0}^{N-1} x[k',l'] g_{l,l'}[k,k']
\end{eqnarray}
\noindent where $g_{l,l'}[k,k']$ represents the \ac{ISI} coefficient relative to the delay-Doppler pair [$k',l'$] seen by sample [$k,l$], whose expression is given by
\begin{equation} \label{eqn:g}
    g_{l,l'}[k,k'] = \sum_{p=0}^{P-1} h'_p \mathbf{\Psi}_{l,l'}^{p}[k,k']
\end{equation}
where the matrix $\mathbf{\Psi}^p \in \mathbb{C}^{NM \times NM}$, is the channel matrix in the delay-Doppler domain with respect to the $p$-th backscattered signal, also referred to as cross-talk matrix. Under some approximations, this matrix can be written as derived in \cite{Gaudio2020} and shown in \eqref{eqn:cross-talk_official}.


\subsection{OTFS-ISAC Input-Output Relationship} \label{sec:OTFS-in-out-rel}

By replacing \eqref{eqn:g} in \eqref{eqn:y_hat_ISAC} and performing a vectorization operation on the transmitted symbol grid of elements $x[k,l]$ and on the received symbol grid of elements $y[k,l]$, the input-output relationship in the delay-Doppler domain can be rewritten in a more compact form as
\begin{equation} \label{eqn:y_hat_ISAC_vect}
    \mathbf{y} = \biggl( \sum_{p=0}^{P-1} h'_p \mathbf{\Psi}^p \biggl) \mathbf{x} + \mathbf{w}
\end{equation}
where $\mathbf{x},\mathbf{y} \in \mathbb{C}^{NM \times 1}$ are the vectors of the transmitted and received symbols, respectively, while $\mathbf{w} \in \mathbb{C}^{NM \times 1}$ represents the \ac{AWGN} vector with mean $\mathbf{0}_{NM}$ and covariance $\mathbf{\Sigma}=\sigma_w^2\mathbf{I}_{NM}$.

Throughout this work, a single point-like target scenario is considered without loss of generality, i.e., $P = 1$, so the index $p$ is dropped from now on. Furthermore, the elements of $\mathbf{x}$ are normalized to unit power, such that $\mathbb{E}\bigl\{\bigl|x[k,l]\bigr|^2\bigr\}=1$, and the complex channel gain is set to $h = 1$ since results are given varying the \ac{SNR}. Therefore, the \ac{SNR} related to the radar part is defined as
\begin{equation}
    \mathrm{SNR}_\mathrm{rad} = \frac{1}{\sigma_w^2}.
\end{equation}
In this setting, \eqref{eqn:y_hat_ISAC_vect} can be rewritten as
\begin{equation} \label{eqn:y_est}
    \mathbf{y} =  h' \mathbf{\Psi} \mathbf{x} + \mathbf{w}
\end{equation}
\noindent where, by considering $h = 1$, the coefficient $h'$, previously defined in Section~\ref{sec:physic-mod}, can be rewritten as $h'=e^{j2\pi f_{\mathrm{D}}\tau}$.


\section{Sensing Parameters Estimation with OTFS and CRLB Calculation} \label{sec:Sensing_estimation}

Considering the received signal in \eqref{eqn:y_est}, and defining $\mathbf{s} = h' \mathbf{\Psi}(\tau,f_{\mathrm{D}}) \mathbf{x}$ the $NM \times 1$ vector of mean values of $\mathbf{y}$, the \ac{ML} estimator for the set of unknown parameters $\mathbf{\boldsymbol{\theta}}~=~(|h'|, \angle{h'},\tau,f_{\mathrm{D}})$, for the considered single-target scenario, is given by
\begin{equation} \label{eqn:ML_est}
    \boldsymbol{\hat{\boldsymbol{\theta}}} = \mathrm{arg} \; \underset{\boldsymbol{\boldsymbol{\theta}} \in \mathbb{R}^4}{\mathrm{max}} \; l(\mathbf{y}|\boldsymbol{\boldsymbol{\theta}},\mathbf{x})
\end{equation}
where $l(\mathbf{y}|\boldsymbol{\theta},\mathbf{x})$ is the log-likelihood function given by \cite{Gaudio2019}
\begin{equation} \label{eqn:likelihood_sub}
    l(\mathbf{y}|\boldsymbol{\boldsymbol{\theta}},\mathbf{x}) = - \biggl(|\mathbf{y}|^2 - \frac{|\mathbf{x}^\mathrm{H} \mathbf{\Psi}(\tau,f_{\mathrm{D}})^\mathrm{H} \mathbf{y}|^2}{\mathbf{x}^\mathrm{H} \mathbf{\Psi}(\tau,f_{\mathrm{D}})^\mathrm{H} \mathbf{\Psi}(\tau,f_{\mathrm{D}}) \mathbf{x}} \biggr).
\end{equation}
Therefore, the estimate ($\hat{\tau},\hat{f}_{\mathrm{D}}$) of $\tau,f_{\mathrm{D}}$, which is directly related to the estimation of the sensing parameters ($r,v$), previously shown in Section~\ref{sec:physic-mod}, can be obtained as
\begin{equation} \label{eqn:argmax}
    (\hat{\tau},\hat{f}_{\mathrm{D}}) = \mathrm{arg} \; \underset{(\tau,f_{\mathrm{D}}) \in \Gamma}{\mathrm{max}} \; \frac{|\mathbf{x}^\mathrm{H} \mathbf{\Psi}(\tau,f_{\mathrm{D}})^\mathrm{H} \mathbf{y}|^2}{\mathbf{x}^\mathrm{H} \mathbf{\Psi}(\tau,f_{\mathrm{D}})^\mathrm{H} \mathbf{\Psi}(\tau,f_{\mathrm{D}}) \mathbf{x}}
\end{equation}
\noindent where $\Gamma$ is a set of delay-Doppler value pairs taken from given delay and Doppler frequency intervals, discretized by steps of $1/(M'\Delta f)$ and $1/(N'T)$, respectively, and assuming $M'~\geq~M$ and $N'~\geq~N$.


The \ac{CRLB} on the estimation of the sensing parameters $\tau$ and $f_{\mathrm{D}}$, considering the received signal in~\eqref{eqn:y_est}, is given by
\begin{equation} \label{eqn:Fisher_var}
    \mathrm{var}[{\hat{\theta}_i}] \geq J_{ii}^{-1}
\end{equation}
\noindent where ${\hat{\theta}_i}$ is the estimate of $\theta_i \in \boldsymbol{\theta}$, representing the $i$-th parameter contained in $\boldsymbol{\theta}$ and $J_{ii}^{-1}$ represents the $i$-th element on the main diagonal of the inverse of the Fisher matrix $\bf J$.

By assuming $s_n$ as the $n$-th element of the expected value vector $\mathbf{s}$, previously defined, and considering the log-likelihood function in \eqref{eqn:likelihood_sub}, the $ij$-th element of the Fisher matrix is defined as \cite{Rife1974}
\begin{equation} 
\label{eqn:Fisher}
    J_{ij} = \frac{2}{\sigma_w^2} \mathfrak{Re} \Biggl\{ \sum_{n=0}^{NM-1} \biggl(\frac{\partial s_n}{\partial \theta_i}\biggr)^\ast \biggl(\frac{\partial s_n}{\partial \theta_j}\biggr) \Biggl\}
\end{equation}
where $s_n$ can be written as an element of a $M \times N$ matrix of indexes $k,l$, as
\begin{equation} \label{eq:s_n}
    s[k,l] = |h'| e^{j \angle h'}\sum_{k'=0}^{M-1}\sum_{l'=0}^{N-1}  \mathbf{\Psi}_{l,l'}[k,k'] x[k',l']
\end{equation}
with $k,k'=0,\dots,M-1$ and $l,l'=0,\dots,N-1$. 

By looking at \eqref{eqn:Fisher} and \eqref{eq:s_n}, it can be noticed that the computation of the \ac{CRLB} reduces to the calculation of the partial derivatives of the cross-talk matrix with respect to the parameters $\tau$ and $f_{\mathrm{D}}$ \cite{Gaudio2020}.


\section{Cross-Talk Matrix Considerations} \label{sec:Cross-talk_computation}

The cross-talk matrix $\mathbf{\Psi} \in \mathbb{C}^{NM \times NM}$ in \eqref{eqn:cross-talk_official}, is the channel matrix in the delay-Doppler domain, containing the information about the sensing parameters. It can be defined as a block matrix since it is composed of $N \times N$ sub-matrices (denoted by indexes $[l,l']$), each of dimension $M \times M$, whose elements are indexed by the pair of indexes $[k,k']$.
As can be seen from \eqref{eqn:cross-talk_official}, a large number of numerical operations is required to compute $\mathbf{\Psi}$. Moreover, according to \eqref{eqn:argmax}, this matrix must be re-computed several times to obtain the estimate of the sensing parameters $\tau$ and $f_{\mathrm{D}}$. For this reason, approaches to reduce the computational complexity of the estimator are very much desirable.

As a starting point for pursuing this objective, the following four matrices are defined
\begin{equation} \label{eqn:Y_1}
    \boldsymbol{\mathsf{Y}}_1 \triangleq \frac{1}{NM} \frac{1-e^{j2\pi(\mathbf{l'}-\mathbf{l}+f_{\mathrm{D}} NT)}}{1-e^{j2\pi\frac{(\mathbf{l'}-\mathbf{l}+f_{\mathrm{D}} NT)}{N}}}
\end{equation}
\begin{equation} \label{eqn:Y_2}
    \boldsymbol{\mathsf{Y}}_2 \triangleq \frac{1}{NM} \frac{e^{-j2\pi(\frac{\mathbf{l'}}{N}+f_{\mathrm{D}} T)}-e^{j2\pi(\mathbf{l'}(1-\frac{1}{N})-\mathbf{l}+f_{\mathrm{D}} T(N-1))}}{1-e^{j2\pi\frac{(\mathbf{l'}-\mathbf{l}+f_{\mathrm{D}} NT)}{N}}}
\end{equation}
\begin{equation} \label{eqn:X_1}
    \boldsymbol{\mathsf{X}}_1 \triangleq \frac{e^{j2\pi f_{\mathrm{D}}\frac{\mathbf{k}'_1}{M\Delta f}}-e^{j2\pi(\mathbf{k}'_1(1+\frac{f_{\mathrm{D}}}{M\Delta f})-\mathbf{k}+\tau M\Delta f)}}{1-e^{j2\pi\frac{\mathbf{k}'_1-\mathbf{k}+\tau M\Delta f}{M}}}
\end{equation}
\begin{equation} \label{eqn:X_2}
    \boldsymbol{\mathsf{X}}_2 \triangleq \frac{e^{j2\pi f_{\mathrm{D}}\frac{\mathbf{k}'_2}{M\Delta f}}-e^{j2\pi(\mathbf{k}'_2(1+\frac{f_{\mathrm{D}}}{M\Delta f})-\mathbf{k}+\tau M\Delta f)}}{1-e^{j2\pi\frac{\mathbf{k}'_2-\mathbf{k}+\tau M\Delta f}{M}}}
\end{equation}
where $\boldsymbol{\mathsf{Y}}_1$, $\boldsymbol{\mathsf{Y}}_2$ are $N \times N$ matrices, while $\boldsymbol{\mathsf{X}}_1$, $\boldsymbol{\mathsf{X}}_2$ have dimension $M \times (M-k_{\tau})$ and $M \times k_{\tau}$, respectively.
The two vectors $\mathbf{l}=[0,\dots,N-1]^\mathrm{T}$ and $\mathbf{l'} = [0,\dots,N-1]$ represent the row and column indexes of $\boldsymbol{\mathsf{Y}}_1$ and $\boldsymbol{\mathsf{Y}}_2$, respectively. Moreover, $\mathbf{k} = [0,\dots,M-1]^\mathrm{T}$ is the row indexes vector of $\boldsymbol{\mathsf{X}}_i$, with $i = 1,2$, while $\mathbf{k}'_1 = [0,\dots,M-1-k_{\tau}]$ and $\mathbf{k}'_2 = [0,\dots,k_{\tau}-1]$ are the column indexes vector of $\boldsymbol{\mathsf{X}}_1$ and $\boldsymbol{\mathsf{X}}_2$, respectively. From these matrices, it is now possible to calculate the cross-talk matrix through the Kronecker products between $\boldsymbol{\mathsf{Y}}_i$ and $\boldsymbol{\mathsf{X}}_i$ matrices with the same indexes and then sum together the resulting matrices thus obtained, as follows
\begin{align} 
    \mathbf{\Psi}_1 &= \boldsymbol{\mathsf{Y}}_1 \otimes [\boldsymbol{\mathsf{X}}_1, \mathbf{0}_{M \times k_\tau}]\nonumber\\
    \mathbf{\Psi}_2 &= \boldsymbol{\mathsf{Y}}_2 \otimes [\mathbf{0}_{M \times (M-k_\tau)},\boldsymbol{\mathsf{X}}_2]\nonumber\\
    \mathbf{\Psi} &= \mathbf{\Psi}_1+\mathbf{\Psi}_2.
\end{align}

To find a method for reducing the computational complexity, first, some considerations about the absolute value of $\boldsymbol{\mathsf{Y}}_1$, $\boldsymbol{\mathsf{Y}}_2$, $\boldsymbol{\mathsf{X}}_1$, $\boldsymbol{\mathsf{X}}_2$ must be made.\footnote{Note that, in this paper, we refer to the absolute value of a matrix as an element-wise modulus operation.} Starting from \eqref{eqn:Y_1}, \eqref{eqn:Y_2}, \eqref{eqn:X_1} and \eqref{eqn:X_2}, it is easy to prove that the absolute value is the same for each of these matrices, and it coincides with the absolute value of the \emph{Dirichlet kernel}.

This function, periodic of $2\pi$ in the $[-\pi,\pi]$ interval, can be conveniently normalized to have periodicity in $[-1/2,1/2]$, as
\begin{equation}
    D_R(x) = \sum_{r=-R}^{R} e^{j2\pi rx} = \frac{\sin((2R+1)\pi x)}{\sin(\pi x)}. 
\end{equation}
With some algebraic manipulations, it can be proved that the absolute value of $\boldsymbol{\mathsf{Y}}_1$ and $\boldsymbol{\mathsf{Y}}_2$
can be expressed as the absolute value of $D_{R}(x)$ apart from a normalization factor $1/(NM)$, when $R=(N-1)/2$, and by defining $x=x_{l,l'} \triangleq x'_{l,l'}/{N}$, with $x'_{l,l'}=l'-l+(f_{\mathrm{D}} NT)$, $\forall l \in \mathbf{l}$, $\forall l' \in \mathbf{l'}$. Similarly, the same expression of the absolute value can be obtained for $\boldsymbol{\mathsf{X}}_1$ and $\boldsymbol{\mathsf{X}}_2$ by assuming $R=(M-1)/2$ and defining $x=x_{k,k'_i} \triangleq x'_{k,k'_i}/{M}$, with $x'_{k,k'_i}=k'_i-k+(\tau M\Delta f)$, $\forall k \in \mathbf{k}$, $\forall k'_{i} \in \mathbf{k'_{i}}$, for $i = 1,2$. More precisely, the aforementioned relationships are
\begin{eqnarray} \label{eqn:D_N} 
    \Bigl|D_{\frac{N-1}{2}}(x'_{l,l'})\Bigr|
    &=&\left|\frac{1-e^{j2\pi x'_{l,l'}}}{1-e^{j2\pi\frac{x'_{l,l'}}{N}}}\right|=|\boldsymbol{\mathsf{Y}}_i[l,l']|
\end{eqnarray}
\begin{eqnarray} \label{eqn:D_M} 
    \Bigl|D_{\frac{M-1}{2}}(x'_{k,k'_i})\Bigr|
    &=&\left|\frac{1-e^{j2\pi x'_{k,k'_i}}}{1-e^{j2\pi\frac{x'_{k,k'_i}}{M}}}\right|=|\boldsymbol{\mathsf{X}}_i[k,k'_{i}]|
\end{eqnarray}
\noindent for $i=1,2$.

The absolute value functions in \eqref{eqn:D_N} and \eqref{eqn:D_M} are periodic of $N$ and $M$ in the intervals $[-N/2,N/2]$ or $[-M/2,M/2]$, respectively. In light of all the previous considerations, the elements of $|\boldsymbol{\mathsf{Y}}_1|$, $|\boldsymbol{\mathsf{Y}}_2|$, $|\boldsymbol{\mathsf{X}}_1|$, and $|\boldsymbol{\mathsf{X}}_2|$ are nothing more than samples of the absolute value of the \emph{Dirichlet kernel}. In particular, each of these samples falls on predictable periodic points with $1$ periodicity, since, by definition, the variable $x'$ can only take value in a specific finite discrete set depending on the related matrix indexes.

The absolute value of the \emph{Dirichlet kernel} can be seen as a periodic function consisting of a main lobe and secondary lobes of lower peak value and decreasing as one moves away from the so-called main lobe.
Due to the geometric property of the considered absolute value function and the fact that the indexes are integers, two samples fall into the main lobe centered at zero, namely in $x'_{l,l'} = x'_{\boldsymbol{\mathsf{Y}},1+} \triangleq \text{frac}(f_{\mathrm{D}}NT)$ and in $x'_{l,l'} = x'_{\boldsymbol{\mathsf{Y}},1-} \triangleq \text{frac}(f_{\mathrm{D}}NT)-1$, for the $|\boldsymbol{\mathsf{Y}}_i|$ matrices, or in $x'_{k, k'_i} = x'_{\boldsymbol{\mathsf{X}},1+} \triangleq \text{frac}(\tau M\Delta f)$ and in $x'_{k,k'_i} = x'_{\boldsymbol{\mathsf{X}},1-} \triangleq \text{frac}(\tau M\Delta f)-1$, for the $|\boldsymbol{\mathsf{X}}_i|$ matrices, with $i=1,2$. Contrariwise, in each secondary lobe only a single sample falls at multiple integers of $x'_{\boldsymbol{\mathsf{Y}},1+}$ or $x'_{\boldsymbol{\mathsf{X}},1+}$ in the positive $x'$ axis and of $x'_{\boldsymbol{\mathsf{Y}},1-}$ or $x'_{\boldsymbol{\mathsf{X}},1-}$ in the negative one, respectively. Thus, it is worth noting that the elements of the four matrices can be seen as periodic samples of the function taken asymmetrically with respect to the y-axis.


\section{Cross-Talk Matrix Approximation} \label{sec:Cross-talk_approx}

By analyzing the cross-talk matrix in \eqref{eqn:cross-talk_official}, it is possible to observe that it is nothing more than a quasi-band matrix since it has only a few elements clustered in bands that have a value in modulus that is not negligible. Moreover, as it will be shown in Section~\ref{sec:Num_Res}, the elements with higher modulus seem to contain most of the information about sensing parameters. Having said that, one way to reduce the computational complexity might be to set the matrix elements with modulus values below a chosen threshold to zero and then treat this matrix as a sparse matrix.

However, given the high dimensionality of the cross-talk matrix and the resulting large number of entries, this approximation strategy would be inefficient from the implementation viewpoint since it would require first computing the entire matrix and then setting some of its elements to zero.

One possible solution to address this problem is not to directly approximate the cross-talk matrix but to approximate the four matrices $|\boldsymbol{\mathsf{Y}}_1|$, $|\boldsymbol{\mathsf{Y}}_2|$, $|\boldsymbol{\mathsf{X}}_1|$ and $|\boldsymbol{\mathsf{X}}_2|$ introduced in Section~\ref{sec:Cross-talk_computation}, which have a reduced dimensionality. In particular, it is possible to select a threshold on each of these matrices, setting it according to the element that falls on one of the lobes of the corresponding \emph{Dirichlet kernel} absolute value function, within one of its periods. These lobes are identified by means of the variable $N_{\mathrm{lobe}}$, with $N_{\mathrm{lobe}} \in [1, \lceil N/2 \rceil]$ for $\boldsymbol{\mathsf{Y}}_i$ functions, and $N_{\mathrm{lobe}} \in [1, \lceil M/2 \rceil]$ for $\boldsymbol{\mathsf{X}}_i$ functions, with $i=1,2$. In particular, $N_{\mathrm{lobe}} = 1$ denotes the main lobe, while values of $N_{\mathrm{lobe}}$ greater than one are used to identify the secondary lobes that are progressively further away from the main lobe, either in the positive or negative $x'$ axis, within a generic period of the considered function. The above approach is motivated by the fact that between the samples on the main lobe and those on one of the secondary lobes, there is a large difference in amplitude, increasing as we move away from the main lobe. 
For this reason, it can be observed that the higher the threshold, the larger the number of elements of the matrix that are left out in this pruning process, leading to a coarser approximation of the matrices \eqref{eqn:Y_1}-\eqref{eqn:X_2}. Going into more detail, it can be observed that the absolute value matrices $|\boldsymbol{\mathsf{Y}}_i|$ and $|\boldsymbol{\mathsf{X}}_i|$, with $i=1,2$, have a well-defined band structure consisting of circularly shifted diagonals. Each of these diagonals is composed of equal elements, whose values are specific samples of the corresponding absolute value function, as stated in Section~\ref{sec:Cross-talk_computation}. The values associated with different diagonals become progressively smaller as one moves away from the main diagonal, which corresponds to the main lobe of the related absolute value function. Furthermore, each diagonal is circularly shifted by a certain factor that depends on the sensing parameters, as will be explained later. Thus, it turns out to be possible to approximate such matrices by band matrices with fewer non-zero diagonals. One way to perform this approximation is to use appropriate matrices that act as masks, having ones where the matrix value must be calculated and zeros otherwise. In particular, these mask matrices are obtained from an identity matrix, by adding to it and then properly circularly shifting a number of super- and sub-diagonals equal to $\lfloor N_{\mathrm{diag}}/2 \rfloor$, where $N_{\mathrm{diag}}$ is the number of non-zero diagonals in the mask matrices, given by
\begin{equation} \label{eqn:N_diag}
    N_{\mathrm{diag}} = 2N_{\mathrm{lobe}}-1,\,\mathrm{with}\,N_{\mathrm{lobe}}\in \mathbb{Z}^+.
\end{equation}
The equality in \eqref{eqn:N_diag} is justified if we consider setting the threshold in correspondence of the element of the four matrices $|\boldsymbol{\mathsf{Y}}_1|$, $|\boldsymbol{\mathsf{Y}}_2|$, $|\boldsymbol{\mathsf{X}}_1|$ and $|\boldsymbol{\mathsf{X}}_2|$, associated with the chosen value of $N_{\mathrm{lobe}}$. Actually, due to the asymmetry of the problem, for a given value of $N_{\mathrm{lobe}}$, there are always two associated elements with different values. However, the equality in \eqref{eqn:N_diag} is still verified if one of the two that is greater in modulus is considered. As already mentioned, the mask matrices actually turn out to be shifted in a manner dependent on the value of delay $\tau$ and Doppler shift $f_{\mathrm{D}}$. In particular, the masks related to $|\boldsymbol{\mathsf{Y}}_1|$ and $|\boldsymbol{\mathsf{Y}}_2|$ turn out to be circularly shifted to the left by $l_{f_{\mathrm{D}}} \triangleq \lceil f_{\mathrm{D}}NT \rceil$ positions, while those associated with $|\boldsymbol{\mathsf{X}}_1|$ and $|\boldsymbol{\mathsf{X}}_2|$ are obtained starting from $M \times M$ mask matrices, by circularly shifting them leftward and downward, respectively, of $k_{\tau} -1$ positions.


\section{Numerical Results} \label{sec:Num_Res}

Numerical simulations are performed to evaluate the impact of the approximation technique of the cross-talk matrix $\mathbf{\Psi}$ introduced in Section~\ref{sec:Cross-talk_approx} on the estimation performance. The analysis is carried out through \ac{RMSE} and square root \ac{CRLB} curves related to the estimation of range and velocity as a function of the \ac{SNR}, and for different values of $N_{\mathrm{lobe}}$, i.e., for different performance/complexity trade-off. System parameters according to the automotive standard IEEE 802.11p are considered. In particular, $f_\mathrm{c}=5.89\,$GHz, $B=10\,$MHz, $M=64$, $N=50$, and $\Delta f=156.25\,$kHz. In addition, a $16$-QAM constellation with unitary mean power symbols and a scenario as the one described in Section~\ref{sec:System_model} are considered.
The \ac{RMSE} is defined as
\begin{equation}
    \mathrm{RMSE}(\hat{z}) = \sqrt{\frac{\sum_{i=0}^{N_{\mathrm{iter}}-1} (\hat{z}_i-z)^2}{N_{\mathrm{iter}}}}
\end{equation}
where $z$ is the true value assumed by a generic parameter to be estimated, and $\hat{z}$ is its estimate, while $N_{\mathrm{iter}}$ is the number of Monte Carlo iterations, here set equal to $1000$.
\begin{figure}[t]
    \centering
\begin{subfigure}[b]{0.42\textwidth}
    \centering
    \includegraphics[width=\textwidth]{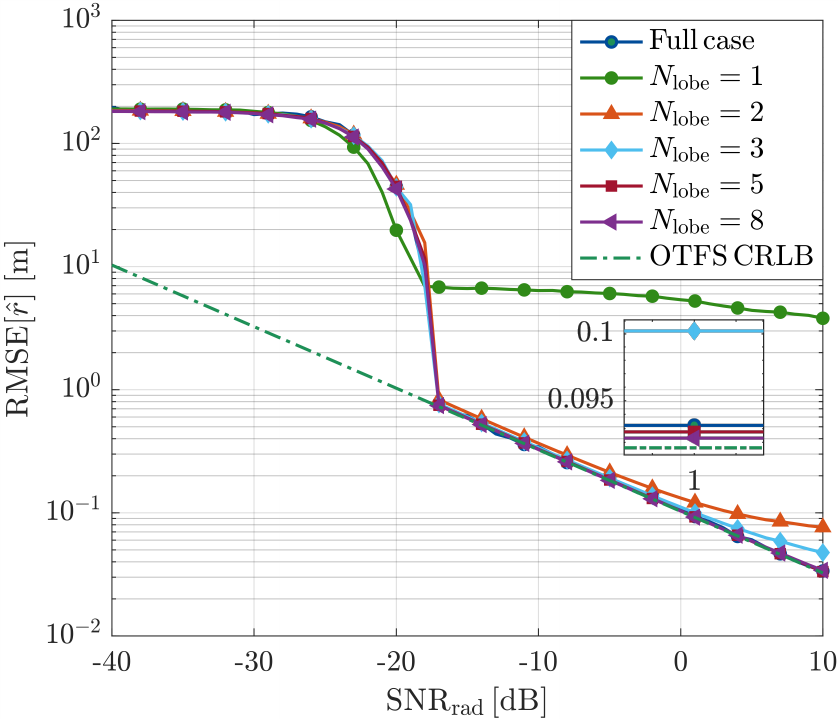}
    \caption{Range estimation RMSE}
    \label{fig:r_RMSE}
\end{subfigure} \qquad \qquad
\begin{subfigure}[b]{0.42\textwidth}
    \centering
    \includegraphics[width=\textwidth]{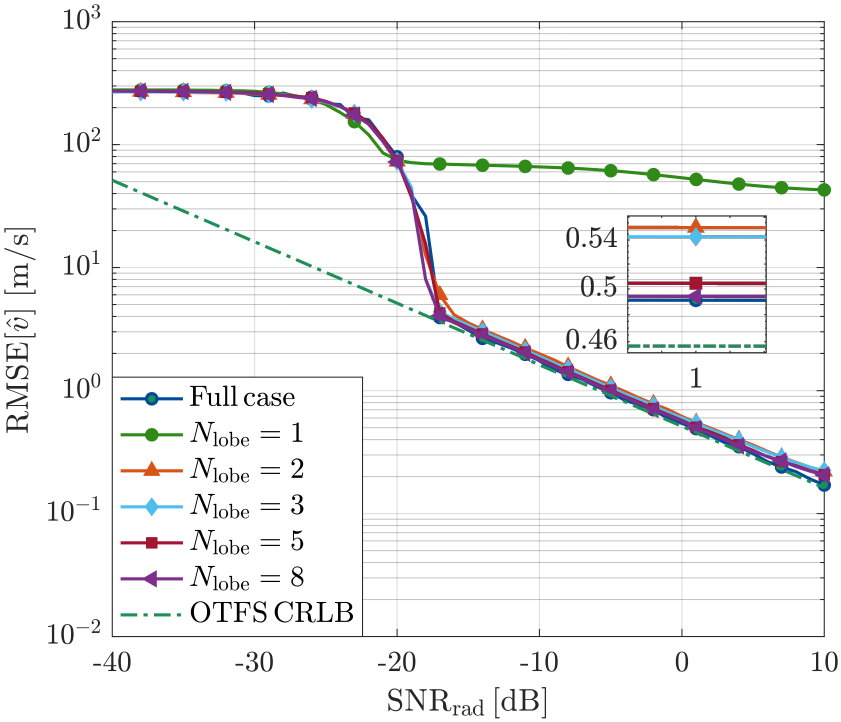}
    \caption{Velocity estimation RMSE}
    \label{fig:vel_RMSE}
\end{subfigure}
\caption{RMSE related to range estimation $\hat{r}$~(a) and velocity estimation $\hat{v}$~(b) for $N_\mathrm{lobe}=1,2,3,5,8$ and for the full case, as a function of SNR, considering $r=20\,$m and $v=80\,$km/h as target parameters.}
\label{fig:RMSE}
\end{figure}
The obtained results are shown in Fig.~\ref{fig:RMSE}. In particular, upon analyzing Fig.~\ref{fig:r_RMSE} and \ref{fig:vel_RMSE}, which show the \ac{RMSE} in the estimation of range and velocity, respectively, it can be observed that the \ac{ISAC} \ac{OTFS}-based system in the case without any approximation, hereinafter referred to as the full case, exhibits good estimation performance up to $\mathrm{\ac{SNR}}=-17$ dB, with the \ac{RMSE} being close to the theoretical bound, for both range and velocity. For what concerns the performance related to the proposed approximation technique, it can be noticed that for high \ac{SNR}, as the value of the variable $N_{\mathrm{lobe}}$ increases, and thus the number of considered elements of the cross-talk matrix, the \ac{RMSE} curves tend to coincide with those related to the full case, i.e., without approximation.

Moreover, even for relatively low values of the variable $N_{\mathrm{lobe}}$, such as $N_{\mathrm{lobe}}=2$, the \ac{RMSE} curves in the approximate case almost match those of the non-approximate case, showing only a marginal degradation with a difference of about $1\,$cm in range and less than $1\,$m/s in velocity in the high-\ac{SNR} regime, but with a significant advantage in terms of computational complexity reduction. In fact, it is possible to verify that when $N_{\mathrm{lobe}}=2$, the computational effort in terms of the number of operations experiences a reduction of nearly $3$ orders of magnitude for each calculation of the cross-talk matrix. Additionally, it is worth noting that for $N_\mathrm{lobe} \geq 5$, the performance becomes nearly identical to that of the full case.
As a result, the proposed approximation criterion is both justified and effective in reducing the complexity of an \ac{ISAC} \ac{OTFS}-based system without significantly compromising the accuracy of sensing parameter estimation when compared to the full case. Consequently, the choice of an appropriate value for $N_{\mathrm{lobe}}$ should be determined by the specific application's requirements, taking into consideration the desired level of estimation accuracy and receiver complexity.


\section{Conclusion} \label{sec:Conclusion}

This work introduced a novel approach to significantly reduce the computational complexity of the \ac{OTFS}-based \ac{ISAC} receiver. The proposed method is based on decomposing the channel matrix into four low-dimensional matrices via algebraic manipulations. Such matrices are then analyzed, and their elements are formulated as samples of the absolute value function of the \emph{Dirichlet kernel} to identify the most informative elements for sensing purposes. Based on this, an analytical criterion, independent of the system parameters, has been proposed to approximate such matrices.\\
Numerical results have shown that the given approximation technique successfully maintains the sensing performance while reducing the computational complexity remarkably, in some cases, of orders of magnitude. A comprehensive analysis of the complexity associated with the discussed approximation technique will be the subject of future work.
This advance can potentially help the adoption of \ac{OTFS} as a modulation for \ac{ISAC} systems, especially for scenarios where devices have limited computational resources and real-time tracking is required.

\bibliographystyle{IEEEtran}
\bibliography{IEEEabrv,bibliography}
\end{document}